# Thermal spin pumping and magnon-phonon-mediated spin-Seebeck effect

K. Uchida,[1, 2, *] T. Ota,[1, 2] H. Adachi,[2, 3] J. Xiao,[4] T. Nonaka,[1, 2] Y. Kajiwara,[1, 2] G. E. W. Bauer,[1, 5] S. Maekawa,[2, 3] and E. Saitoh[1, 2, 3, 6]

[1]*Institute for Materials Research, Tohoku University, Sendai 980-8577, Japan*
[2]*CREST, Japan Science and Technology Agency, Sanbancho, Tokyo 102-0075, Japan*
[3]*Advanced Science Research Center, Japan Atomic Energy Agency, Tokai 319-1195, Japan*
[4]*Department of Physics and State Key Laboratory of Surface Physics, Fudan University, Shanghai 200433, China*
[5]*Kavli Institute of NanoScience, Delft University of Technology, 2628 CJ Delft, The Netherlands*
[6]*WPI Advanced Institute for Materials Research, Tohoku University, Sendai 980-8577, Japan*

The spin-Seebeck effect (SSE) in ferromagnetic metals and insulators has been investigated systematically by means of the inverse spin-Hall effect (ISHE) in paramagnetic metals. The SSE generates a spin voltage as a result of a temperature gradient in a ferromagnet, which injects a spin current into an attached paramagnetic metal. In the paramagnet, this spin current is converted into an electric field due to the ISHE, enabling the electric detection of the SSE. The observation of the SSE is performed in longitudinal and transverse configurations consisting of a ferromagnet/paramagnet hybrid structure, where thermally generated spin currents flowing parallel and perpendicular to the temperature gradient are detected, respectively. Our results explain the SSE in terms of a two-step process: (1) the temperature gradient creates a non-equilibrium state in the ferromagnet governed by both magnon and phonon propagations and (2) the non-equilibrium between magnons in the ferromagnet and electrons in the paramagnet at the contact interface leads to "thermal spin pumping" and the ISHE signal. The non-equilibrium state of metallic magnets (e.g. $Ni_{81}Fe_{19}$) under a temperature gradient is governed mainly by the phonons in the sample and the substrate, while in insulating magnets (e.g. $Y_3Fe_5O_{12}$) both magnon and phonon propagations appear to be important. The phonon-mediated non-equilibrium that drives the thermal spin pumping is confirmed also by temperature-dependent measurements, giving rise to a giant enhancement of the SSE signals at low temperatures.

## I. INTRODUCTION

The Seebeck effect or generation of an electric voltage by a temperature gradient in conductors was discovered by T. J. Seebeck in the 1820s [see Fig. 1(a)].[1] In contrast, the spin-Seebeck effect (SSE) stands for the generation of a 'spin voltage' as a result of a temperature gradient in magnetic materials,[2–17] where spin voltage refers to potential to drive a spin current,[18–21] i.e. a flow of electron spin-angular momentum into an attached conductor [see Fig. 1(b)]. The SSE is important in spintronics[22–27] and spin caloritronics,[28–33] since it enables simple and versatile generation of spin currents from heat. In 2008, we observed the SSE in a ferromagnetic metal $Ni_{81}Fe_{19}$ film[2] by means of a spin-detection technique based on the inverse spin-Hall effect[34–46] (ISHE) in a Pt film. In 2010, Jaworski *et al.* also observed this phenomenon in a ferromagnetic semiconductor GaMnAs at low temperatures (below Curie temperature of GaMnAs) by the same method.[6,12] In the same year, we revealed that the SSE appears even in magnetic insulators, such as $Y_3Fe_5O_{12}$ (YIG),[7] $La Y_2Fe_5O_{12}$ (La:YIG),[5] and $(Mn,Zn)Fe_2O_4$.[9] The SSE was observed as well in the half-metallic Heusler compound $Co_2MnSi$.[13] These observations indicate that the SSE is a universal phenomenon in magnetic materials [see Figs. 1(c) and 1(d)].

The discovery of the SSE in magnetic insulators provides a crucial piece of information for understanding the physics of the SSE. The conventional Seebeck effect requires itinerant charge carriers, or conduction electrons, and therefore exists only in metals and semiconductors [see Fig. 1(c)]. It appeared natural to assume that the same held for the SSE. In fact, originally the SSE was phenomenologically formulated in terms of thermal excitation of conduction electrons.[2] However, the observation of the SSE in insulators upsets this conventional assumption; conduction electrons are not necessary for the SSE. This is the direct evidence that the spin voltage generated by the SSE is associated with magnetic dynamics [see Fig. 1(b)]. Based on this idea, various theoretical models have been proposed.[4,8,10,11]

In this paper, we systematically investigate the SSE in ferromagnetic metals and ferrimagnetic insulators by means of the ISHE in paramagnetic metals. This paper is organized as follows. In Sec. II, we explain sample configurations and measurement setups for the experiments. In Sec. III, we provide the details of the measurement procedures and report the observation of the SSE in various sample systems. The basic theoretical concepts of the SSE are reviewed in Sec. IV. The last Sec. V is devoted to conclusions.

## II. SAMPLE CONFIGURATION AND MEASUREMENT MECHANISM

The observation of the SSE exploits the ISHE in two different device structures: one is a *longitudinal* configuration,[7,9] in which a spin current *parallel* to a temperature gradient is measured. This structure is the simplest and most straightforward one, but applicable only for magnetic insulators, as mentioned below. The other setup is a *transverse* configuration,[2,3,5,6,12,13] in which a spin current flowing *perpendicular* to a temperature gradient is measured. The transverse configuration has a more complicated device structure (hence is more difficult to measure) than the longitudinal one, but it has been used to measure the SSE in various materials. The first observation of the SSE in $Ni_{81}Fe_{19}$ films was reported in the transverse configuration.[2]

Figure 2(a) shows a schematic illustration of the longitudinal SSE device. The device structure is very simple,

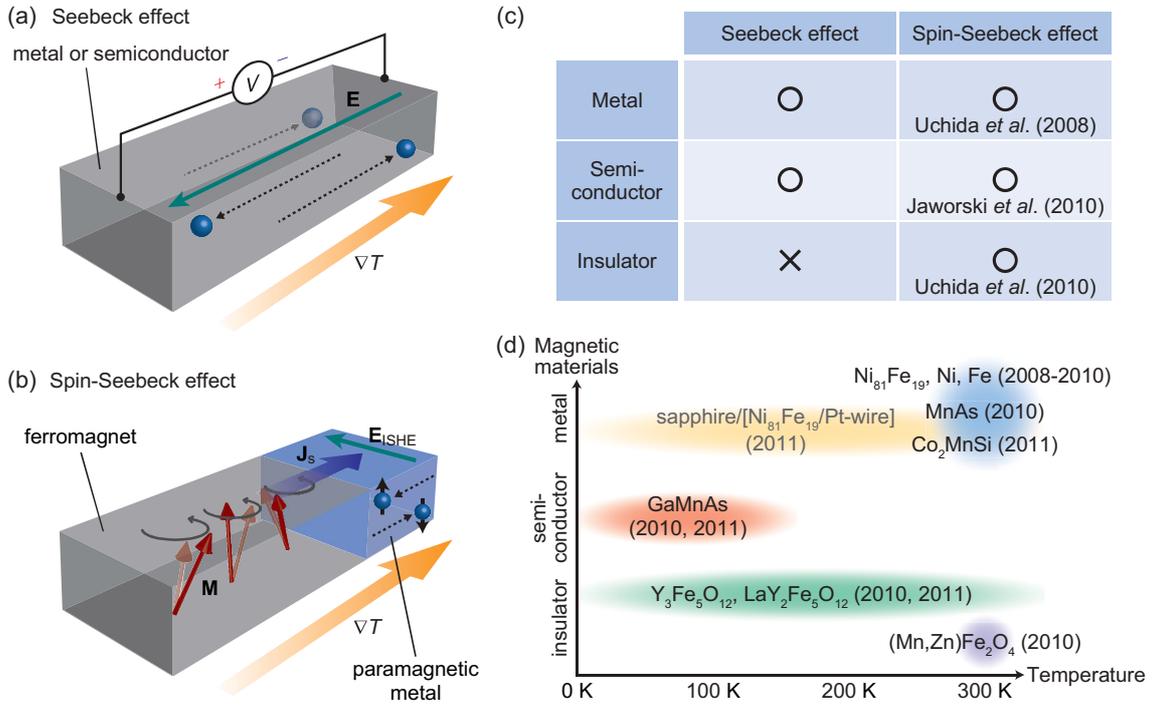

FIG. 1: (a) A schematic illustration of the conventional Seebeck effect. When a temperature gradient $\nabla T$ is applied to a conductor, an electric field $\mathbf{E}$ (electric voltage $V$) is generated along the $\nabla T$ direction. (b) A schematic illustration of the spin-Seebeck effect (SSE). When $\nabla T$ is applied to a ferromagnet, a spin voltage is generated via magnetization ($\mathbf{M}$) dynamics, which pumps a spin current $\mathbf{J}_s$ into an attached paramagnetic metal. In the paramagnetic metal, this spin current is converted into an electric field $\mathbf{E}_{\text{ISHE}}$ due to the inverse spin-Hall effect (ISHE). (c) Difference between the Seebeck effect and the SSE. The SSE appears even in insulators. (d) Experimental reports on the SSEs.

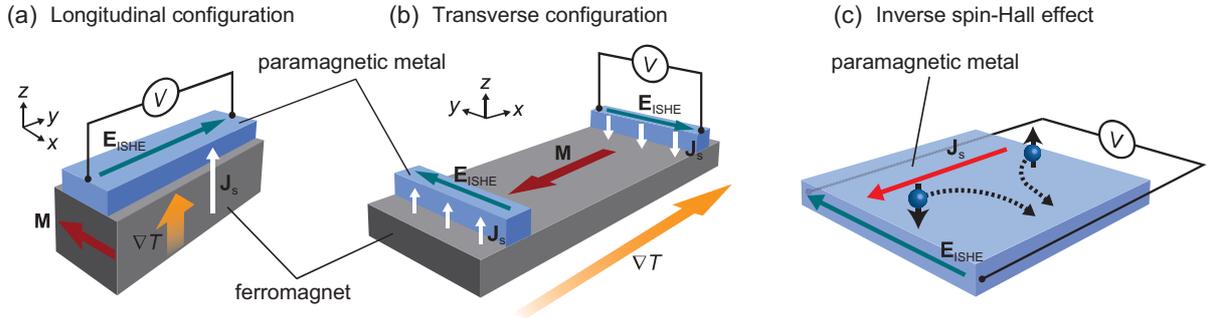

FIG. 2: [(a), (b)] Schematic illustrations of the longitudinal configuration (a) and the transverse configuration (b) for measuring the SSE. (c) A schematic illustration of the ISHE.

which consists of a ferromagnet (F) slab covered with a paramagnetic metal (PM) film. When a temperature gradient $\nabla T$ is applied over the F layer perpendicular to the F/PM interface ($z$ direction), a spin voltage is thermally generated and injects a spin current with the spatial direction $\mathbf{J}_s$ along the $\nabla T$ direction and the spin-polarization vector $\boldsymbol{\sigma}$ parallel to the magnetization $\mathbf{M}$ of F, into the PM [see Fig. 2(a)]. The spin current is converted into an electric field $\mathbf{E}_{\text{ISHE}}$ by the ISHE. When $\mathbf{M}$ is along the $x$ direction, $\mathbf{E}_{\text{ISHE}}$ is generated in the PM along the $y$ direction according to the relation

$$\mathbf{E}_{\text{ISHE}} = \frac{\theta_{\text{SH}}\rho}{A}\left(\frac{2|e|}{\hbar}\right)\mathbf{J}_s \times \boldsymbol{\sigma}, \tag{1}$$

where $\theta_{\text{SH}}$, $\rho$, and $A$ are the spin-Hall angle of PM, the electric resistivity of PM, and the contact area between F and PM, respectively [see Figs. 2(a) and 2(c)].[47] $\theta_{\text{SH}}$ is especially large in noble metals with strong spin-orbit interaction, such as Pt. Therefore, by measuring $\mathbf{E}_{\text{ISHE}}$ in the PM film, one can detect the longitudinal SSE electrically. Here we note that, if metal slabs were as F, the ISHE signal was not only suppressed significantly by short-circuit currents[44] in F but was also contaminated by the anomalous Nernst voltage.[48] By using an insulator such as YIG, these artifacts are eliminated.

Figure 2(b) shows a schematic illustration of the transverse SSE device, which consists of a rectangular-shaped F with one or several PM wires attached on its top surface. The typical length of the F layer along the $x$ direction is ~ 10 mm, much longer than the conventional spin-diffusion length.[49] In the transverse configuration, $\nabla T$ is applied along the $x$ direction. In order to generate an ISHE voltage, the F layer has to be magnetized along the $\nabla T$ direction [see Eq. (1) and Fig. 2(b) and note that $\mathbf{J}_s$ is parallel to the $z$ direction also in the transverse configuration]. Therefore, the anomalous Nernst effect in the F layer vanishes since $\nabla T$ and $\mathbf{M}$ are collinear, enabling an unperturbed detection of the transverse SSE in various magnetic materials. The characteristic property of the transverse SSE is the sign reversal of the thermally generated spin voltage between the lower- and higher-temperature ends of the F layer.[2,3,5,6,12,13] Therefore, the sign of the resultant ISHE voltage ($\mathbf{E}_{\mathrm{ISHE}}$) is also reversed [see Fig. 2(b)], which is direct evidence for the transverse SSE.

## III. ELECTRIC DETECTION OF THE SPIN-SEEBECK EFFECT BY THE INVERSE SPIN-HALL EFFECT

### A. Longitudinal spin-Seebeck effect

#### 1. Measurement system

In Fig. 3, we show a photograph and a schematic illustration of a measurement system used for the longitudinal SSE experiments in the present study. The longitudinal F/PM sample, illustrated in Fig. 2(a), is sandwiched between two Cu plates; the upper Cu plate is attached to a heat bath of which the temperature is controlled by a closed-cycle helium refrigerator and the lower one is placed on the top surface of a Peltier thermoelectric module. Here, the bottom surface of the Peltier module is thermally connected to the heat bath. By applying an electric current to the Peltier module, the temperature of the lower Cu plate is increased or decreased, and a temperature gradient is generated in the F/PM sample along the $z$ direction [see Fig. 3(b)]. We measured the temperature difference between the upper and lower Cu plates with two T-type thermocouples. As shown in Fig. 3(c), to measure an electric voltage $V$ between the ends of the PM layer of the longitudinal SSE sample, tungsten needles were attached to the ends of the sample by using a micro-probing system [note that the length of the sample (6 mm) is slightly longer than the width of the upper Cu plate (5 mm)]. To avoid electrical contact of the upper Cu plate with the PM layer, a silicone-rubber sheet was inserted between them. Since the thickness of the silicone rubber is very thin ($\sim 0.1$ mm), the sample is thermally well connected to the upper Cu block. An external magnetic field $\mathbf{H}$ (with the magnitude $H$) was applied to the YIG/Pt samples in the $x$-$y$ plane at an angle $\theta$ to the $y$ direction.

#### 2. Observation of the longitudinal spin-Seebeck effect in ferrimagnetic insulator/paramagnetic metal systems

First, we show data on the longitudinal SSE in ferrimagnetic insulator/Pt samples. Figure 4(a) shows a photograph of the sample system used in this experiment. The sample consists of a single-crystalline or polycrystalline YIG slab and a Pt film sputtered on a well-polished YIG surface. The length, width, and thickness of the YIG slab (Pt film) are 6 mm (6 mm), 2 mm (0.5 mm), and 1 mm (15 nm), respectively.

Figure 4(c) shows $V$ between the ends of the Pt layer in the single-crystalline YIG/Pt sample as a function of the temperature difference $\Delta T$ at $H = 1\,\mathrm{kOe}$. When $\mathbf{H}$ is applied along the $x$ direction ($\theta = 90°$), the magnitude of $V$ is observed to be proportional to $\Delta T$. The sign of the $V$ signal for finite values of $\Delta T$ is clearly reversed by reversing the temperature gradient. Since YIG is an insulator, thermoelectric phenomena in itinerant magnets, such as the conventional Seebeck and Nernst effects, do not exist at all. As also shown in Fig. 4(c), the $V$ signal disappears when $\mathbf{H}$ is along the $y$ direction ($\theta = 0$), a situation consistent with the symmetry of the ISHE induced by the longitudinal SSE [see Eq. (1) and Fig. 4(b)].

To confirm the origin of this signal, we measured the magnetic field ($H$) dependence of $V$ in the same YIG/Pt system. We found that the sign of $V$ is reversed by reversing $\mathbf{H}$ when $\theta = 90°$ and $|H| > 500\,\mathrm{Oe}$, indicating that the $V$ signal is affected by the magnetization direction of the YIG layer [see Fig. 4(d)]. The $V$ signal disappears when the Pt layer is replaced by a paramagnetic metal film with weak spin-orbit interaction, such as Cu. This behavior supports the expected longitudinal SSE scenario.

Figure 4(e) shows the $\Delta T$ and $H$ dependences of $V$ in the polycrystalline YIG/Pt sample, which also exhibits the longitudinal SSE. As shown in Ref. 10, the longitudinal SSE was observed even in sintered polycrystalline insulating magnets, such as $(\mathrm{Mn,Zn})\mathrm{Fe_2O_4}$.

The critical difference between the SSEs in single-crystalline and polycrystalline samples becomes apparent in the temperature dependence of $V$. Figure 5 shows $V/\Delta T$ ($\equiv S_{\mathrm{LSSE}}$) as a function of the temperature $T$ in the single-crystalline and polycrystalline YIG/Pt samples at $H = 1$ kOe. In the whole temperature range (4.2–290 K), clear SSE signals were observed in both the samples for $\mathbf{H}$ along the $\theta = 90°$ direction. Notable is that, in the single-crystalline YIG/Pt sample, the magnitude of $V/\Delta T$ is dramatically enhanced around $T = 50$ K, while the $V$ signal in the polycrystalline YIG/Pt sample does not exhibit strong $T$ dependence (see Fig. 5). We found that this $V$-peak position coincides with the temperature at which thermal conductivity of the single-crystalline YIG becomes its maximum due to the increase of the phonon lifetime,[50] suggesting the importance of the phonons in creating the non-equilibrium state that drives the spin current in the Pt contact.[8,12,15] This phonon-mediated contribution was observed also in the transverse configuration, as shown in Sec. III B.

### B. Transverse spin-Seebeck effect

#### 1. Measurement system

In Figs. 6(a) and 6(b), we present detailed information on the measurement system used for the transverse SSE experiments consisting of a Cu block attached to a heat bath and a 0.2-mm-thick Cu plate thermally isolated from the block by a 2-mm-thick bakelite board. A strain gauge of 120 $\Omega$ and two T-type thermocouples are respectively attached on the top of the Cu plate and on the ends of a dummy substrate bridged between the Cu block and the Cu plate [see Figs. 6(a) and 6(b)]. The transverse F/PM sample, illustrated in Fig. 2(b), is fixed near the dummy substrate. By applying an electric current to the strain gauge, the temperature of the Cu plate increases, and the temperature difference $\Delta T$ is generated between the ends of the sample along the $x$ direction [see Fig. 6(b)]. We measured an electric voltage difference $V$ between the ends of the PM wires in the transverse F/PM sample under the temperature gradient by using the aforementioned micro-probing system with applying $\mathbf{H}$ along the $x$ direction, except when collecting magnetic-filed-angle-dependent data.

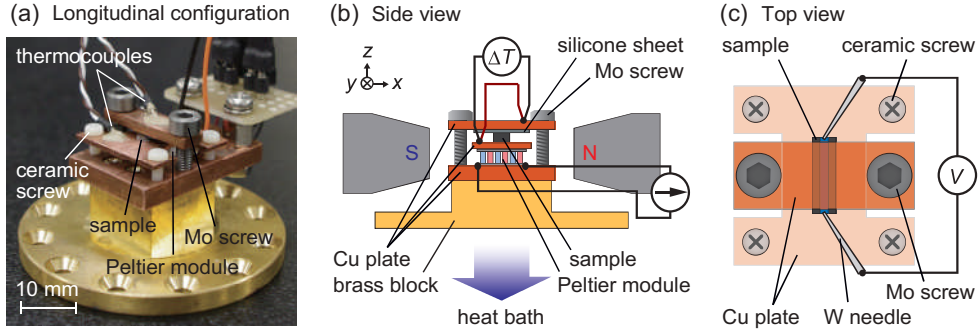

FIG. 3: A photograph (a) and a schematic illustration [(b): side view, (c): top view] of the measurement system for the longitudinal SSE experiments. The Cu plate above the sample ('upper Cu plate' in the main text) is thermally well connected to a heat bath through thick (M3) molybdenum (Mo) screws with high thermal conductivity ($\sim 140\,\mathrm{Wm^{-1}K^{-1}}$), while the the Cu plate just below the sample ('lower Cu plate' in the main text) is thermally insulated from the heat bath by thin (M2) ceramic screws with low thermal conductivity ($< 1\,\mathrm{Wm^{-1}K^{-1}}$). Since the diameter of the tip of the tungsten (W) needles is very small ($\sim 10\ \mu$m), the heat flow from the needles to the sample is negligibly small.

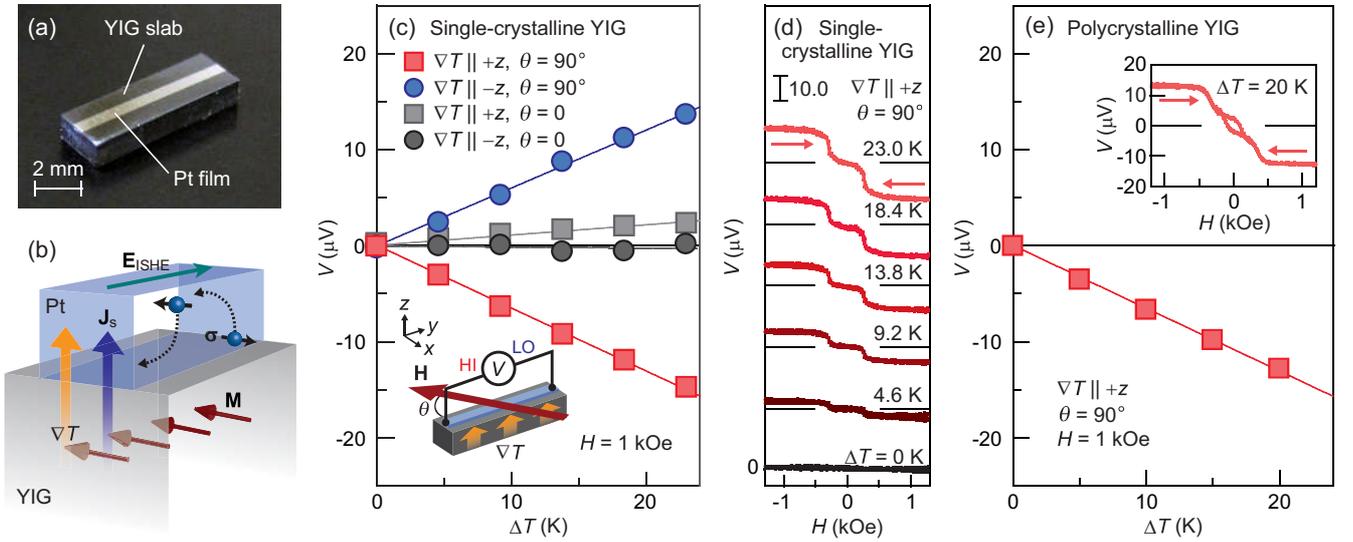

FIG. 4: (a) A photograph of the YIG/Pt sample in the longitudinal configuration. (b) A schematic illustration of the longitudinal SSE and the ISHE in the YIG/Pt sample. (c) $\Delta T$ dependence of $V$ in the single-crystalline YIG/Pt sample at $H = 1\,\mathrm{kOe}$, measured when $\nabla T$ was applied along the $+z$ (upward) or $-z$ (downward) direction. The magnetic field $\mathbf{H}$ was applied along the $x$ direction ($\theta = 90°$) or the $y$ direction ($\theta = 0$). (d) $H$ dependence of $V$ in the single-crystalline YIG/Pt sample for various values of $\Delta T$ at $\theta = 90°$, measured when $\nabla T$ was along the $+z$ direction. (e) $\Delta T$ dependence of $V$ in the polycrystalline YIG/Pt sample at $H = 1\,\mathrm{kOe}$ and $\theta = 90°$, measured when $\nabla T$ was along the $+z$ direction. The inset to (e) shows the $H$ dependence of $V$ in the polycrystalline YIG/Pt sample at $\Delta T = 20\,\mathrm{K}$. All the data shown in this figure were measured at room temperature.

We checked that the system shown in Fig. 6(a) can generate a uniform temperature gradient before each measurement of the transverse SSE. For example, Fig. 6(c) shows a temperature ($T$) image and profile along the $x$ direction of an insulating La:YIG/Pt sample used for the experiments in Sec. III B 2, measured with an infrared camera. The $T$ image shows that the temperature distribution in the La:YIG/Pt sample has a linear profile along the $x$ direction and there are no temperature variations along the $y$ direction. We also confirmed a uniform temperature gradient in a metallic $\mathrm{Ni_{81}Fe_{19}}$ film fabricated on a sapphire substrate, which is used for the experiments in Sec. III B 3, by measuring thermoelectric voltage in the $\mathrm{Ni_{81}Fe_{19}}$ film along the $x$ direction [see Fig. 6(d)]. The linear distribution of the thermoelectric voltage observed here bears witness to a uniform temper-ature gradient in the $Ni_{81}Fe_{19}$ film, which is owing to the almost same thermal conductivities of the $Ni_{81}Fe_{19}$ film and the sapphire substrate at room temperature.[51] Here we note that, if thermal conductivity mismatch between a film and a substrate was large, the SSE measurements were disturbed by a perpendicular temperature gradient that induces parasitic Nernst signals.[13,17,53]

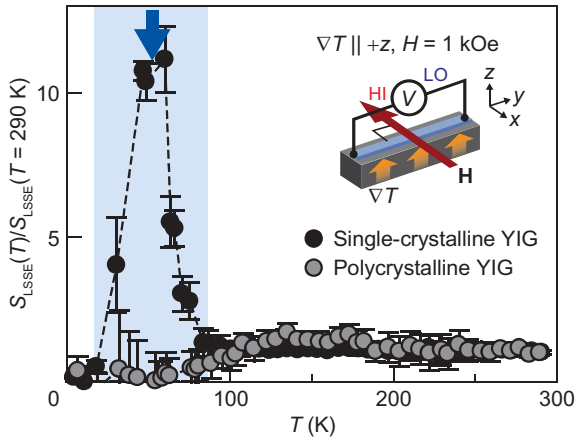

FIG. 5: $T$ dependence of $S_{LSSE}(T)/S_{LSSE}(T = 290$ K) in the single-crystalline and polycrystalline YIG/Pt samples at $H = 1$ kOe, measured when $\nabla T$ and **H** were applied along the $+z$ and $x$ directions, respectively. $S_{LSSE}(T)$ denotes the ISHE voltage induced by the longitudinal SSE per unit temperature difference: $S_{LSSE}(T) = V(T)/\Delta T$. The vertical axis is normalized by $S_{LSSE}(T)$ at 290 K.

### 2. Observation of the transverse spin-Seebeck effect in ferrimagnetic insulator/paramagnetic metal systems

In this subsection, we discuss results on the transverse SSE in ferrimagnetic insulator/Pt systems. The sample consists of a La:YIG film with Pt wires attached to the La:YIG surface [see Figs. 7(a) and 7(b)]. A single-crystalline La:YIG (111) film with the thickness of 3.9 $\mu$m was grown on a paramagnetic $Gd_3Ga_5O_{12}$ (GGG) (111) substrate by liquid phase epitaxy. By substituting a part of yttrium in YIG by lanthanum, the lattice matching between the fabricated garnet film and the substrate is improved. Here, the surface of the La:YIG layer has an 8×4 mm² rectangular shape. Two (and later more) 15-nm-thick Pt wires were then sputtered in an Ar atmosphere on the top of the La:YIG film. The distance of the Pt wires from the center of the La:YIG layer is 2.8 mm. The length and width of the Pt wires are 4 mm and 0.1 mm, respectively. The resistance between the Pt wires is much greater than 10 GΩ, indicating that the wires are electrically well insulated.

Figure 7(c) shows $V$ between the ends of the Pt wires placed near the lower- and higher-temperature ends of the La:YIG layer as a function of $\Delta T$ at $H = 100$ Oe. The magnitude of $V$ is proportional to $\Delta T$ in both Pt wires. Notably, the sign of $V$ for finite values of $\Delta T$ is clearly reversed between the lower- and higher-temperature ends of the sample. This sign reversal of $V$ is a characteristic behavior of the ISHE voltage induced by the transverse SSE[2,3,5,6,12,13] (see Sec. IV). As shown in Fig. 7(d), the sign of the $V$ signal at each end of the sample is reversed by reversing **H**. We also measured $V$ by changing the angle $\theta$ of the in-plane magnetic field to the $x$ direction and the $V$ signal was found to vanish when $\theta = 90°$, a situation consistent with Eq. (1) [see Fig. 7(e)]. This $V$ signal disappears when the Pt wires are replaced by Cu wires with weak spin-orbit interaction [see Fig. 7(f)]. We checked that the signal also disappears in a La:YIG/SiO₂/Pt system,[14] in which the La:YIG and Pt layers are separated by a thin (10 nm) film of insulating SiO₂, as well as in a GGG/Pt system, in which the Pt wire is directly fabricated on the GGG substrate, indicating that the direct contact between La:YIG and Pt is essential for generating the $V$ signal. An extrinsic proximity effect or induced ferromagnetism in the Pt layers can be excluded because of the sign change of $V$ between the ends of the La:YIG/Pt sample. All the data shown above confirm that the observed $V$ signal is due entirely to the transverse SSE in the La:YIG/Pt sample.

Up to now, we discussed the spin voltage generated near the ends of the La:YIG layer. In contrast, Fig. 8(a) shows $V/\Delta T$ as a function of the position of the Pt wire ($x_{Pt}$) for various values of the temperature $T$ in the La:YIG film with nine Pt wires attached. Since each Pt wires are electrically well insulated, one can investigate the spatial distribution of $V$ systematically for the same sample. Note that this setup is impossible for metallic ferromagnet systems since short-circuit currents in the ferromagnet disturb the spatial profile of the ISHE voltage. In the La:YIG/Pt sample, the $V$ signal clearly increases (decreases) for $x_{Pt} > 0$ ($x_{Pt} < 0$) and disappears at the center of the sample, indicating that the spin voltage generated from a uniform temperature gradient varies almost linearly along the $\nabla T$ direction in the present millimeter-sized La:YIG film (see Sec. IV). We found that, above 200 K, $V$ varies almost linearly with respect to $x_{Pt}$, while the $x_{Pt}$ dependence of $V$ deviates from the linear function below 150 K; the magnitude of $V$ decays exponentially over several millimeters distance from both ends of the La:YIG layer. This result indicates the existence of a characteristic length of the SSE, which is fundamentally different from conventional spin-diffusion lengths. We estimated the characteristic length $\lambda$ of the SSE in the La:YIG film by fitting the experimental data in Fig. 8(a) by the hyperbolic sine function $\xi \sinh(x/\lambda)$, where $\xi$ is a constant. As shown in Fig. 8(b), $\lambda$ reaches several millimeters and decreases with decreasing $T$. According to the magnon-based mechanism (see Sec. IV),[4,5,10,11] this $T$ dependence of $\lambda$ can be qualitatively explained by the increase of magnetic damping in La:YIG at low temperatures,[52] which disturbs long-range magnon propagation.

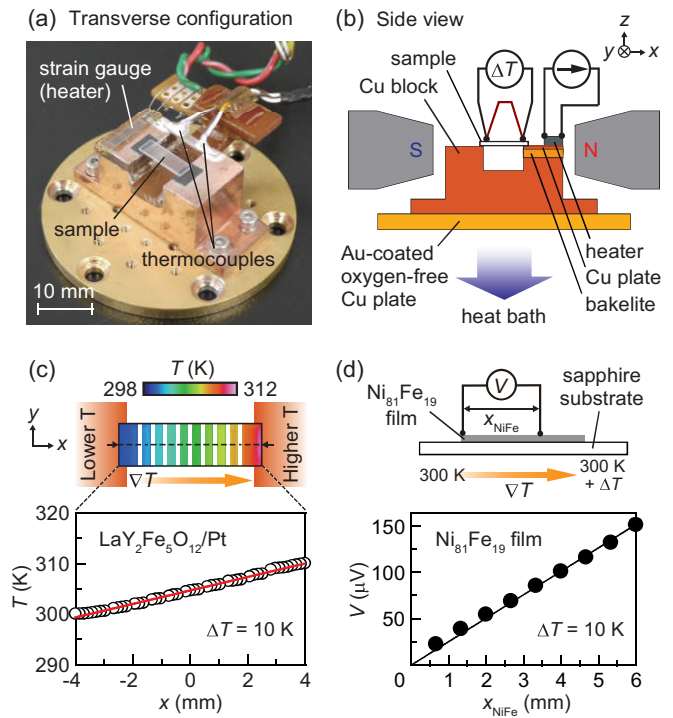

FIG. 6: [(a), (b)] A photograph (a) and a schematic illustration (b) of the measurement system for the transverse SSE experiments. (c) A temperature image and profile along the $x$ direction of a La:YIG/Pt sample for $\Delta T = 10$ K, measured with an infrared camera. Temperatures of the metallic films cannot be measured due to very low infrared emittance. (d) $x_{NiFe}$ dependence of the thermoelectric voltage in a $Ni_{81}Fe_{19}$ film placed on a sapphire substrate at $\Delta T = 10$ K, where $x_{NiFe}$ denotes the distance between the electrodes attached to the $Ni_{81}Fe_{19}$ film. The position of one electrode was fixed at the lower-temperature end of the film.

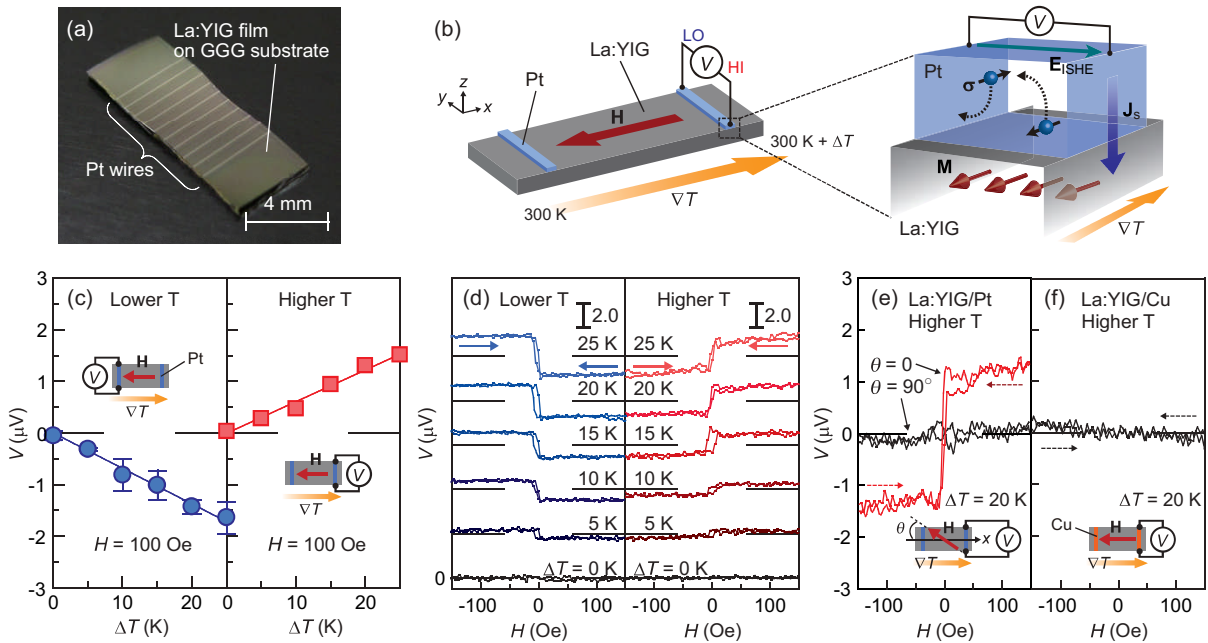

FIG. 7: (a) A photograph of the La:YIG/Pt sample in the transverse configuration. The experimental data in this figure were measured for a La:YIG film with two Pt wires, although the sample in this photograph has ten wires. (b) A schematic illustration of the transverse SSE and the ISHE in the La:YIG/Pt sample. (c) $\Delta T$ dependence of $V$ in the La:YIG/Pt sample at $H = 100$ Oe, measured when the Pt wires were attached near the lower-temperature (300 K) and higher-temperature (300 K+$\Delta T$) ends of the La:YIG layer. (d) $H$ dependence of $V$ in the La:YIG/Pt sample for various values of $\Delta T$. (e) $H$ dependence of $V$ in the La:YIG/Pt sample at $\Delta T = 20$ K, measured when $\mathbf{H}$ was applied at an angle $\theta$ to the $x$ direction. (f) $H$ dependence of $V$ in a La:YIG/Cu sample at $\Delta T = 20$ K, measured when $\mathbf{H}$ was along the $x$ direction. The experimental data shown in (e) and (f) were measured at the higher-temperature end of the sample.

Next, we focus on the temperature dependence of the magnitude of the SSE signal in the transverse La:YIG/Pt sample. Figure 8(c) shows $V/\Delta T$ as a function of $T$ at $H = 100$ Oe. When the Pt wires are attached near the ends of the La:YIG layer, the enhancement of the SSE signal was observed also in this transverse configuration at $T = 50$ K, which is explained by the model calculation based on the phonon-mediated process.[8]

### 3. Observation of the transverse spin-Seebeck effect in ferromagnetic metal/paramagnetic metal systems

In this subsection, we show the experimental results on the transverse SSE in ferromagnetic metal/Pt samples. We prepared a transverse SSE device in which the insulating La:YIG layer is replaced with a ferromagnetic metal $Ni_{81}Fe_{19}$. The sample consists of a 20-nm-thick $Ni_{81}Fe_{19}$ film with a 10-nm-thick Pt wire attached to the end of the $Ni_{81}Fe_{19}$. Here, note that each $Ni_{81}Fe_{19}$/Pt sample has *only one* Pt wire in order to reduce short-circuit currents [see Fig. 9(b)];[44] if two or more Pt wires were attached to a same $Ni_{81}Fe_{19}$ film, the wires were electrically connected, complicating the investigation of the transverse SSE signals. The $Ni_{81}Fe_{19}$ layer was deposited by electron-beam evaporation in a high vacuum on a single-crystalline sapphire (0001) substrate, of which the thermal conductivity is almost the same as that of $Ni_{81}Fe_{19}$ at room temperature.[51] Then, the Pt layer was sputtered in an Ar atmosphere on the end of the $Ni_{81}Fe_{19}$ film [see Fig. 9(a)]. Immediately before the sputtering, the surface of the $Ni_{81}Fe_{19}$ layer was cleaned by Ar-ion etching. The surface of the $Ni_{81}Fe_{19}$ layer has an 6×4 $mm^2$ rectangular shape. The length and width of the Pt wire are 4 mm and 0.1 mm, respectively.

Figures 9(c) and 9(d) respectively show the $\Delta T$ and $H$ dependences of $V$ in the $Ni_{81}Fe_{19}$/Pt sample at $H = 100$ Oe, measured when the Pt wire was placed on the lower- and higher-temperature ends of the $Ni_{81}Fe_{19}$ film. The measurements were performed using the identical sample [see Fig. 9(b)]. The $V$ behaviors observed in the $Ni_{81}Fe_{19}$/Pt sample are almost the same as those of the transverse SSE in the La:YIG/Pt samples. Notably, in our experimental setup, a temperature gradient in the $Ni_{81}Fe_{19}$ film *perpendicular* to the film plane (along the $z$ direction)[53] is negligibly small, since the $V$ signal disappears in a plain $Ni_{81}Fe_{19}$ film to which no Pt wire is attached [see Fig. 9(e)]. Therefore, the $V$ signal observed in the $Ni_{81}Fe_{19}$/Pt sample is not associated with the conventional Seebeck and/or Nernst effects in an unconnected $Ni_{81}Fe_{19}$ film. As mentioned above, the uniform in-plane temperature gradient in our $Ni_{81}Fe_{19}$/Pt sample is realized owing to the almost same thermal conductivities of the $Ni_{81}Fe_{19}$ film and the sapphire substrate.[51]

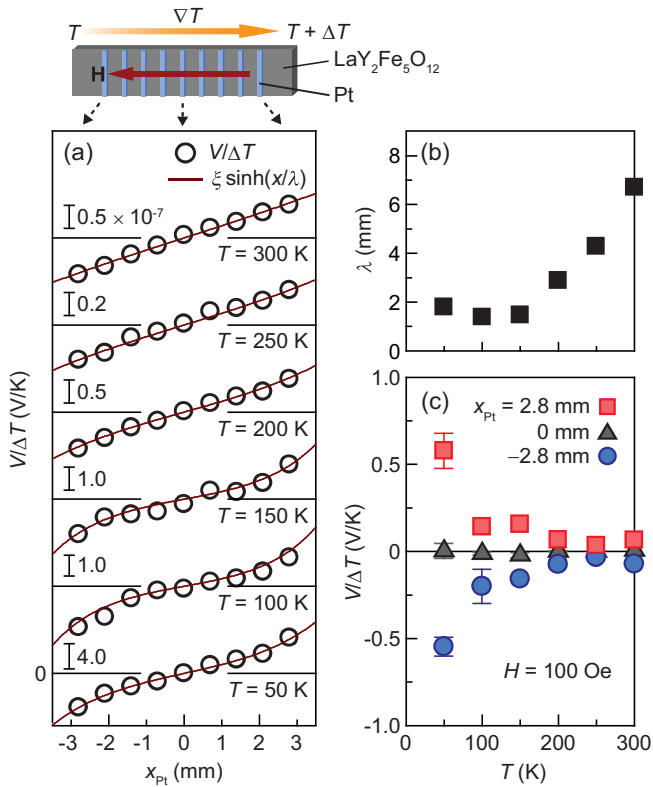

FIG. 8: (a) Dependence of $V/\Delta T$ on $x_{Pt}$, the displacement of the Pt wire from the center of the La:YIG layer along the $x$ direction, in the La:YIG/Pt sample for various values of $T$ at $H = 100$ Oe (solid circles). The temperatures of the lower- and higher-temperature ends of the sample were stabilized to $T$ and $T + \Delta T$, respectively. The solid curves are the fitting results using a hyperbolic sine function $\xi \sinh(x/\lambda)$, where $\xi$ and $\lambda$ are adjustable parameters. (b) $T$ dependence of $\lambda$. (c) $T$ dependence of $V/\Delta T$ at $H = 100$ Oe, measured when the Pt wires were placed at $x_{Pt} = 2.8$ mm, 0 mm, and $-2.8$ mm.

To investigate the spatial distribution of the thermally generated spin voltage, we measured $V$ using four separate $Ni_{81}Fe_{19}$/Pt samples in which a Pt wire is placed at different positions on the $Ni_{81}Fe_{19}$ film. The Pt wire in each $Ni_{81}Fe_{19}$/Pt sample is placed perpendicular to the $x$ ($\nabla T$) direction. In Fig. 9(f), we show $V$ as a function of $x_{Pt}$ for various values of $\Delta T$ at $H = 100$ Oe.

The observed spatial profile of the SSE signal in the $Ni_{81}Fe_{19}$/Pt samples was found to be quite similar to that in the La:YIG/Pt samples [compare Figs. 8(a) and 9(f)]. However, this long-range spin-voltage profile in ferromagnetic metals cannot be explained by the magnon-based mechanism since the strong magnetic damping in ferromagnetic metals drastically suppresses the magnon propagation length. Obviously, this long-range nature of the SSE in the $Ni_{81}Fe_{19}$/Pt sample cannot also be reproduced by the conventional spin-diffusion equation because this equation expresses that spin voltage (spin accumulation) decays within the spin-diffusion-length scale, which is very short (submicrometer scale).[54] A key to solve this mystery is provided by the experiments shown in Sec. III B 4.

#### 4. Acoustic spin-Seebeck effect

The biggest unsolved issue of the SSE is that the thermally generated spin voltage appears over a millimeter scale even in ferromagnetic *metals* in the transverse configuration. In this subsection, we show that the long-range feature of the transverse SSE in ferromagnetic metals can be explained by ballistic phonon propagation.

To demonstrate the essential role of phonons in the metal SSE, we prepared a 20-nm-thick $Ni_{81}Fe_{19}$/10-nm-thick Pt bilayer wire placed on a single-crystalline sapphire (0001) substrate [see Fig. 10(a)]. The substrate is of $10 \times 3$-$mm^2$ rectangular shape. The length and width of the $Ni_{81}Fe_{19}$/Pt wire are 3 mm and 0.1 mm, respectively. The distance of the $Ni_{81}Fe_{19}$/Pt wire from the center of the sapphire substrate is 3.2 mm. Here, this wire is completely isolated both electrically and magnetically since there are no electric and spin carriers in the sapphire. Owing to this structure, only phonons can pass through the substrate [see Fig. 10(b)]. Therefore, if the SSE appears even in this sapphire/[$Ni_{81}Fe_{19}$/Pt-wire] structure, it will become a conclusive proof for the existence of phonon-mediated mechanisms in the SSE in ferromagnetic metals. We measured the electric voltage difference $V$ between the ends of the Pt layer of the sapphire/[$Ni_{81}Fe_{19}$/Pt-wire] sample with applying $\nabla T$ and **H** at 300 Oe along the $x$ direction [see Fig. 10(a)].

Figure 10(c) shows $V$ as a function of $\Delta T$ in the sapphire/[$Ni_{81}Fe_{19}$/Pt-wire] sample, measured when the $Ni_{81}Fe_{19}$/Pt wire was placed near the lower- and higher-temperature ends of the substrate. Surprisingly, a clear SSE signal appears even in this structure; this $V$ behavior is consistent with the aforementioned feature of the transverse SSE, which cannot be explained by conventional thermoelectric effects since the temperature gradient in the sapphire substrate is uniform (confirmed with an infrared camera)[15] and the $Ni_{81}Fe_{19}$/Pt wire is isolated both electrically and magnetically. In fact, the $V$ signal disappears both when **H** is applied along the $y$ direction ($\theta = 90°$) and when the Pt layer is replaced with a Cu film [see Fig. 10(c)], consistent with Eq. (1).

Next, to show how the $V$ signal varies with changing the $Ni_{81}Fe_{19}$/Pt-wire position on the sapphire, we attached ten $Ni_{81}Fe_{19}$/Pt wires on the sapphire substrate and measured $V$ in the wires. These wires are separated from each other far enough to cut electric and magnetostatic coupling between the wires. As shown in Fig. 10(d), in the sapphire/[$Ni_{81}Fe_{19}$/Pt-wire array] sample, $V$ varies almost linearly with the position of the $Ni_{81}Fe_{19}$/Pt wire; all the results confirm that the transverse SSE appears even in the isolated $Ni_{81}Fe_{19}$/Pt wire on the single-crystalline sapphire substrate.

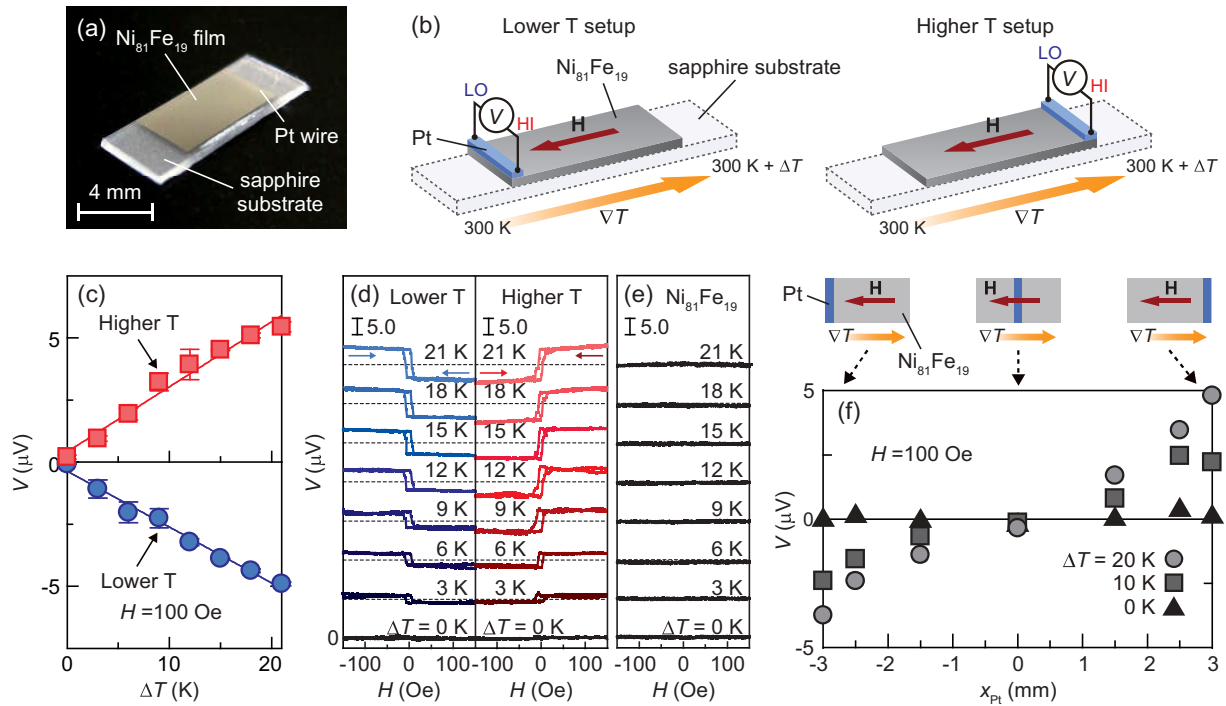

FIG. 9: (a) A photograph of the $Ni_{81}Fe_{19}$/Pt sample in the transverse configuration. (b) A schematic illustration of the measurement setup for the transverse SSE in the $Ni_{81}Fe_{19}$/Pt sample. (c) $\Delta T$ dependence of $V$ in the $Ni_{81}Fe_{19}$/Pt sample at $H = 100$ Oe, measured when the Pt wire was attached to the lower-temperature (300 K) and higher-temperature ($300\,K + \Delta T$) ends of the $Ni_{81}Fe_{19}$ layer. (d) $H$ dependence of $V$ in the $Ni_{81}Fe_{19}$/Pt sample for various values of $\Delta T$. (e) $H$ dependence of $V$ in a plain $Ni_{81}Fe_{19}$ film for various values of $\Delta T$. (f) $x_{Pt}$ dependence of $V$ in the $Ni_{81}Fe_{19}$/Pt samples for various values of $\Delta T$ at $H = 100$ Oe.

The only possible mechanism of the SSE in this sapphire/[$Ni_{81}Fe_{19}$/Pt-wire] structure is phonon-mediated spin dynamics. Since phonons can pass through even an insulating substrate, the distribution function of magnons in the $Ni_{81}Fe_{19}$ wire is modulated by the non-equilibrium phonons through the magnon-phonon interaction. This modulation activates the thermal spin pumping into the Pt layer [see Fig. 10(b)].[15] Notable is that, since phonons with low frequencies (less than 20 THz: the thermal-phonon-densest frequency at 300 K) exhibit very long propagation length, magnons in the $Ni_{81}Fe_{19}$ wire can "feel" substrate temperature at positions far away from the wire, yielding the close-to-linear dependence and sign reversal of the SSE signal between the lower- and higher-temperature regions of the sample. A model calculation based on the linear-response theory[15] suggests that the SSE signal in the present setup is proportional to the phonon lifetime in the substrate and a parameter reflecting the acoustic-impedance-matching condition[55] between the substrate and the ferromagnetic metal layer. We confirmed this acoustic mechanism by measuring $V$ in a glass/[$Ni_{81}Fe_{19}$/Pt-wire array] sample where the single-crystalline sapphire substrate is replaced with a same-sized silica-glass substrate [see Fig. 10(d)]. In the glass/[$Ni_{81}Fe_{19}$/Pt-wire array] sample, the same temperature gradient is formed [compare Figs. 10(e) and 10(f)]. Nevertheless, the $V$ signal was observed to disappear in the glass/[$Ni_{81}Fe_{19}$/Pt-wire array] sample, a situation attributed to the short phonon lifetime in the glass substrate and the acoustic-impedance mismatch between the glass and the $Ni_{81}Fe_{19}$ layers[56] [see Fig. 10(d)].

To further buttress the importance of the phonons in the substrate, we measured the temperature ($T$) dependence of the SSE signal in the sapphire/[$Ni_{81}Fe_{19}$/Pt-wire] sample. Figure 11 shows $V/\Delta T$ between the ends of the Pt layer as a function of $T$, measured when the $Ni_{81}Fe_{19}$/Pt wire was placed near the higher-temperature end of the sapphire substrate. The clear SSE signals appear in all the temperature range. Notably, the magnitude of $V/\Delta T$ is strongly enhanced around $T = 40$ K, a situation similar to the results shown in Figs. 5 and 8(c). This $V$ enhancement also provides a crucial evidence that the SSE signal in the sapphire/[$Ni_{81}Fe_{19}$/Pt-wire] sample is dominated by the phonon-mediated process through the sapphire substrate, since the $V$-peak structure at low temperatures corresponds to the increase of the phonon lifetime, i.e. thermal conductivity, in the sapphire substrate.[57]

The phonon-mediated SSE observed here is responsible for the long-range feature of the transverse SSE observed in the $Ni_{81}Fe_{19}$-film/Pt-wire sample [see Fig. 9(f)]; in the conventional $Ni_{81}Fe_{19}$/Pt sample, the phonon-mediated process through the $Ni_{81}Fe_{19}$ film can explain the long-range spin-voltage distribution in the $Ni_{81}Fe_{19}$ (see Fig. 13). The SSE-induced ISHE voltage in the Ni$_{81}$Fe$_{19}$-film/Pt-wire sample [Fig. 9(c)] is one order of magnitude greater than that observed in the sapphire/[Ni$_{81}$Fe$_{19}$/Pt-wire] sample [Fig. 10(c)], a situation explained by the fact that the path of phonons in the present sapphire/[Ni$_{81}$Fe$_{19}$/Pt-wire] sample is limited to the substrate; since phonon heat currents do not flow into the bulk of the Ni$_{81}$Fe$_{19}$ wire due to the heat balance condition, phonons interact with magnons only near the sapphire/Ni$_{81}$Fe$_{19}$ interface, while on the other hand, in the Ni$_{81}$Fe$_{19}$-film/Pt-wire sample, phonons interact with magnons in the bulk of the Ni$_{81}$Fe$_{19}$ (see Fig. 13). Phonon mediation also explains the so-called scratch experiment by Jaworski *et al.*,[6] who showed that the SSE signal in their GaMnAs/Pt sample does not change before and after scratching a ferromagnetic GaMnAs layer. This is the evidence that the SSE in their sample is dominated by the phonon propagation through the substrate.[8]

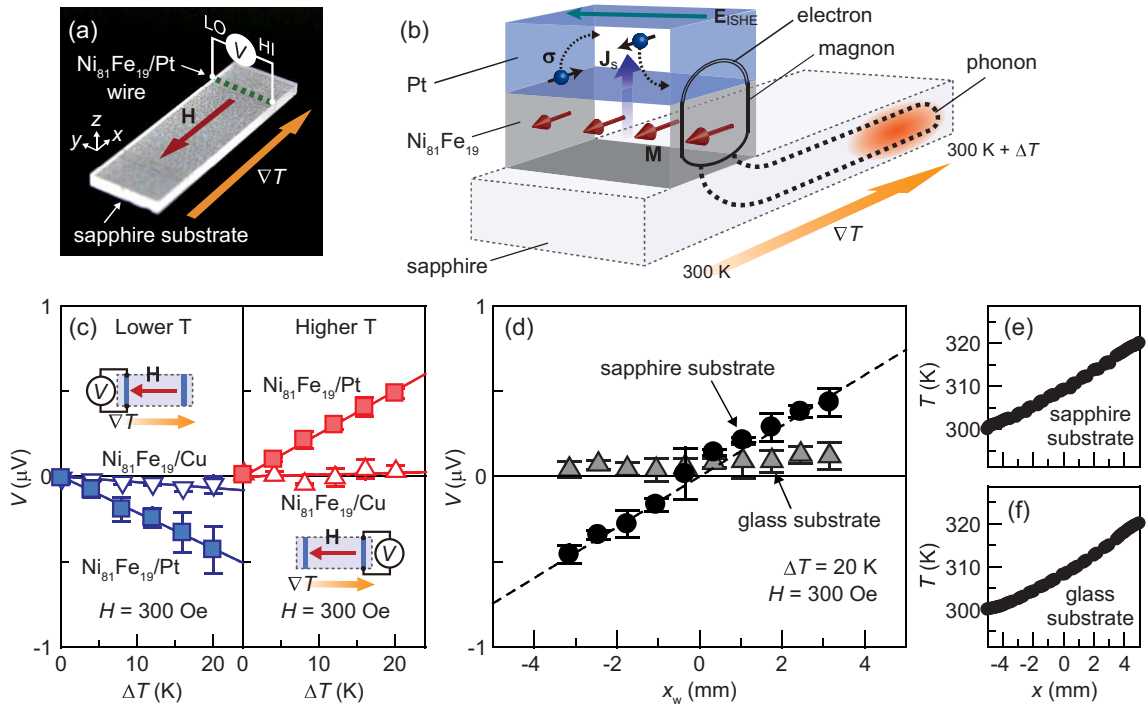

FIG. 10: (a) A photograph of the sapphire/[Ni$_{81}$Fe$_{19}$/Pt-wire] sample for the measurement of the acoustic SSE. (b) A schematic illustration of the acoustic SSE and the ISHE in the sapphire/[Ni$_{81}$Fe$_{19}$/Pt-wire] sample. The double lines, bold lines, and dotted lines represent electron spin-density propagators, magnon propagators, and phonon propagators (see also Fig. 13), respectively. (c) $\Delta T$ dependence of $V$ in the sapphire/[Ni$_{81}$Fe$_{19}$/Pt-wire] and sapphire/[Ni$_{81}$Fe$_{19}$/Cu-wire] samples at $H = 300$ Oe, measured when the Ni$_{81}$Fe$_{19}$/Pt and Ni$_{81}$Fe$_{19}$/Cu wires were respectively placed near the lower-temperature (300 K) and higher-temperature (300 K$+\Delta T$) ends of the sapphire substrate. (d) $x_w$ dependence of $V$ in the sapphire/[Ni$_{81}$Fe$_{19}$/Pt-wire array] and glass/[Ni$_{81}$Fe$_{19}$/Pt-wire array] samples at $\Delta T = 20$ K and $H = 300$ Oe, where $x_w$ denotes the displacement of the Ni$_{81}$Fe$_{19}$/Pt wire from the center of the substrate along the $x$ direction. [(e), (f)] Temperature profiles along the $x$ direction of the sapphire/[Ni$_{81}$Fe$_{19}$/Pt-wire array] (e) and glass/[Ni$_{81}$Fe$_{19}$/Pt-wire array] (f) samples at $\Delta T = 20$ K.

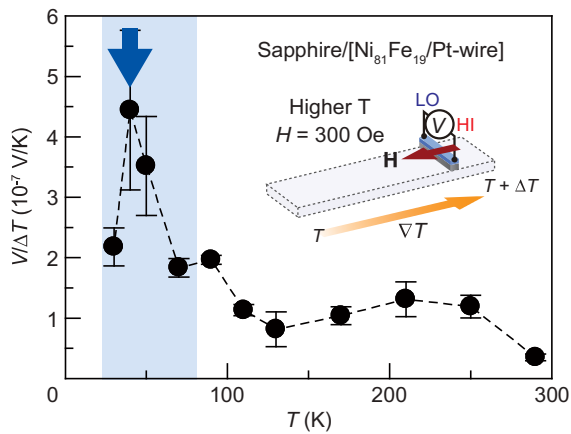

FIG. 11: $T$ dependence of $V/\Delta T$ in the sapphire/[Ni$_{81}$Fe$_{19}$/Pt-wire] sample at $H = 300$ Oe, measured when the Ni$_{81}$Fe$_{19}$/Pt wire was placed near the higher-temperature end of the sapphire substrate.

## IV. THEORETICAL CONCEPT OF THE SPIN-SEEBECK EFFECT

We now present a qualitative model of the mechanism for the SSE. Since the SSE appears even in magnetic insulators, it cannot fully be expressed in terms of thermal excitation of conduction electrons. The SSE in insulators cannot be explained also by equilibrium spin pumping,[21,34,37,40–46,58–61] since the average spin-pumping current from thermally fluctuating magnetic moments in a ferromagnet (F) is exactly canceled by the thermal (Johnson-Nyquist) spin-current noise[62,63] from an attached paramagnetic metal (PM). Therefore, the observed spin voltage requires us to introduce a non-equilibrium state between magnetic moments in F and electrons in PM. Microscopic theories for the SSE have been proposed by Xiao et al.[4] and by Adachi et al.[8,10,15] by means of scattering and linear-response theories, respectively. In this section, we review basic concepts of their calculations.

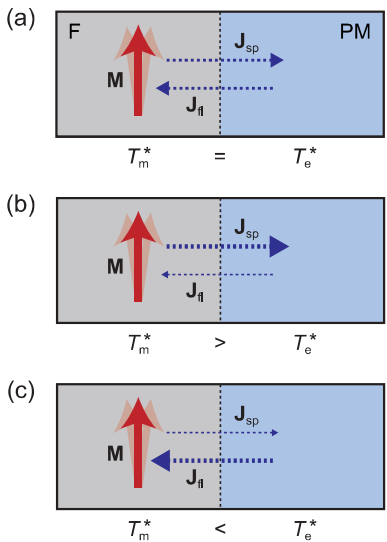

(a)

(b)

(c)

FIG. 12: Mechanism of the spin-current generation by the SSE at the ferromagnet (F)/paramagnetic metal (PM) interface. $\mathbf{J}_{\mathrm{sp}}$ and $\mathbf{J}_{\mathrm{fl}}$ denote the thermal spin-pumping current from F to PM proportional to an effective magnon temperature $T_{\mathrm{m}}^{*}$ in F and the Johnson-Nyquist spin-current noise from PM to F proportional to an effective electron temperature $T_{\mathrm{e}}^{*}$ in PM, respectively. The dimension of $\mathbf{J}_{\mathrm{sp}}$ and $\mathbf{J}_{\mathrm{fl}}$ is Joule, i.e. the flow of $\hbar/2$ per unit time.

A thermally excited state in the SSEs at an F/PM interface can be described in terms of an effective magnon temperature $T_{\mathrm{m}}^{*}$ in F and an electron temperature $T_{\mathrm{e}}^{*}$ in PM, which are allowed to differ. The ISHE signal derived from the net spin current is proportional to $T_{\mathrm{m}}^{*} - T_{\mathrm{e}}^{*}$, as shown below. The effective temperatures are related to thermal fluctuations through the fluctuation-dissipation theorem. The fluctuations of the magnetization $\mathbf{m}$ at the F/PM interface are excited by a random magnetic field $\mathbf{h} = \Sigma_{j}\mathbf{h}^{(j)}$ ($j = 0, 1, ...$), which represents the thermal disturbance from various sources (such as lattice, contacts etc.). If we respectively denote $T^{(j)}$ and $\alpha^{(j)}$ as the temperature and the damping of dissipative source $j$, due to the fluctuation-dissipation theorem, the random field $\mathbf{h}$ satisfies the following equal-position time-correlation function:

$$\left\langle h_i^{(j)}\left(t\right) h_{i'}^{(j')}\left(t'\right)\right\rangle = \left(\frac{2k_{\mathrm{B}}T^{(j)}\alpha^{(j)}}{\gamma M_{\mathrm{s}}V_{\mathrm{a}}}\right)\delta_{jj'}\delta_{ii'}\delta(t-t'), \quad (2)$$

where $i$, $i' = x$, $y$, $z$ (position index), $k_{\mathrm{B}}$ is the Boltzmann constant, $\gamma$ the gyromagnetic ratio, $M_{\mathrm{s}}$ the saturation magnetization, $\alpha = \alpha^{(0)} + \alpha^{(1)} + ...$ the effective damping parameter, $T_{\mathrm{m}}^{*} = [\alpha^{(0)}T^{(0)}+\alpha^{(1)}T^{(1)}+...]/\alpha$ the effective magnon temperature, and $V_{\mathrm{a}}$ the magnetic coherence volume in F, which depends on the magnon temperature and the spin-wave stiffness constant $D$. When the dissipative sources 0 and 1 are the F lattice and the PM contact respectively, then $T^{(0)}$, $\alpha^{(0)}$, $T^{(1)}$, and $\alpha^{(1)} = \gamma\hbar g_{\mathrm{r}}/4\pi M_{\mathrm{s}}V_{\mathrm{a}}$ represent the bulk lattice temperature, the bulk Gilbert damping parameter, the electron temperature in the PM contact ($T^{(1)} = T_{\mathrm{e}}^{*}$), and the damping enhancement due to the spin pumping with $g_{\mathrm{r}}$ being the real part of the mixing conductance for the F/PM interface. The net thermal spin current across the F/PM interface is given by the sum of a fluctuating thermal spin-pumping current $\mathbf{J}_{\mathrm{sp}}$ from F to PM proportional to $T_{\mathrm{m}}^{*}$ and a Johnson-Nyquist spin-current noise $\mathbf{J}_{\mathrm{fl}}$ from PM to F proportional to $T_{\mathrm{e}}^{*}$:[4,62,63]

$$\mathbf{J}_{\mathrm{s}} = \mathbf{J}_{\mathrm{sp}} + \mathbf{J}_{\mathrm{fl}} = \frac{M_{\mathrm{s}}V_{\mathrm{a}}}{\gamma}\left[\alpha^{(1)}\mathbf{m}\times\dot{\mathbf{m}} + \gamma\mathbf{m}\times\mathbf{h}^{(1)}\right]. \quad (3)$$

The DC component along the magnetization equilibrium direction ($x$ direction) reduces to

$$J_{\mathrm{s}} \equiv \langle\mathbf{J}_{\mathrm{s}}\rangle_x = 2\alpha^{(1)}k_{\mathrm{B}}(T_{\mathrm{m}}^{*} - T_{\mathrm{e}}^{*}). \quad (4)$$

Therefore, when $T_{\mathrm{m}}^{*} > T_{\mathrm{e}}^{*}$ ($T_{\mathrm{m}}^{*} < T_{\mathrm{e}}^{*}$), spin currents are injected from F (PM) into PM (F) [see Figs. 12(b) and 12(c)]. At equilibrium ($T_{\mathrm{m}}^{*} = T_{\mathrm{e}}^{*}$), no spin current is generated at the F interface since the spin-pumping current, $\mathbf{J}_{\mathrm{sp}}$, is canceled out by the spin-current noise, $\mathbf{J}_{\mathrm{fl}}$ [see Fig. 12(a)]. Equations (1) and (4) indicate that the magnitude and sign of the ISHE voltage induced by the SSE are determined by those of $T_{\mathrm{m}}^{*} - T_{\mathrm{e}}^{*}$ at the F/PM under a temperature gradient. As demonstrated by the previous studies, this effective temperature difference is induced by magnon-mediated[4,5,10,11] and phonon-mediated[8,12,15] processes, where $T_{\mathrm{m}}^{*}$ in F and/or $T_{\mathrm{e}}^{*}$ in PM are modulated by magnons and phonons propagating through F under a temperature gradient, respectively. Due to these processes, magnons in F and/or electrons in PM in a lower-temperature (higher-temperature) region feel temperature information in a higher-temperature (lower-temperature) region. Therefore, the resultant $T_{\mathrm{m}}^{*}$ and/or $T_{\mathrm{e}}^{*}$ in the lower-temperature (higher-temperature) region increases (decreases).

An important clue to the effective temperature distribution was provided by Sanders and Walton in 1977.[64] They discussed the effective magnon-temperature ($T_{\mathrm{m}}^{*}$) and phonon-temperature ($T_{\mathrm{p}}^{*}$) distributions in a magnetic insulator, especially YIG, under a uniform temperature gradient and solved a simple heat-rate equation of the coupled magnon-phonon system under a situation similar to the transverse SSE configuration. The solution of the heat-rate equation yields a hyperbolic sine $T_{\mathrm{m}}^{*} - T_{\mathrm{p}}^{*}$ profile with a decay length $\lambda_{\mathrm{m}}$. In a magnetic insulator with weak magnetic damping, such as YIG and La:YIG, a rapid equilibration of magnons is prevented and the resulting $\lambda_{\mathrm{m}}$ was shown to reach several millimeters; when the sample length is comparable to $\lambda_{\mathrm{m}}$, a hyperbolic sine function becomes close to a linear function. Making use of these results and assuming that $T_{\mathrm{p}}^{*}$ in F is equal to $T_{\mathrm{e}}^{*}$ in an attached PM, Xiao et al. formulated the magnon-mediated SSE and their calculation quantitatively reproduces the magnitude and spatial distribution of the SSE-induced ISHE signal observed in the transverse La:YIG/Pt system.[4,5] The magnon-mediated SSE was formulated also by means of a many-body technique using non-equilibrium Green's functions by Adachi et al.[10] and numerical calculation based on a stochastic Landau-Lifshitz-Gilbert equation by Ohe et al.[11] It is important to point out that, in the latter two approaches, the concept of local heat bath temperature plays a crucial role, and the definition and the interpretation of the effective magnon and phonon temperatures are different from those of Sanders and Walton.

A mechanism according to which the effective temperature difference $T_m^* - T_e^*$ is generated by non-equilibrium phonons was formulated by Adachi *et al.*[8,15] using a linear-response theory. This process can explain the following experimental results: (1) the giant enhancement of the SSE signal at low temperatures[8,12] [Figs. 5 and 8(c)], (2) the long-range spin-voltage distribution in ferromagnetic metals [Fig. 9(f)], and (3) the acoustic SSE [Figs. 10 and 11], which cannot be explained by the Sanders-Walton mechanism of diffuse heat transport.

The dominant contributions to the SSEs in various configurations can be represented by the Feynman diagrams shown in Fig. 13,[5,7,8,10,15] which assist intuitive understanding of the microscopic processes of the SSEs. In the transverse configuration, since heat currents flow only in the F layer due to the heat balance condition, the modulation of $T_m^*$ in F through the magnon- and/or phonon-mediated processes gives a dominant contribution to the SSE. As discussed above, the phonon-mediated process is essential in the metallic $\mathrm{Ni_{81}Fe_{19}/Pt}$ samples, but on the other hand both the magnon- and phonon-mediated processes can contribute to the SSE in the insulating YIG/Pt and La:YIG/Pt samples. In contrast, in the case of the longitudinal SSE in the YIG/Pt samples, we found that the dominant contribution comes from the excitation of conduction electrons (modulation of $T_e^*$) in the Pt layer through the phonon-mediated process due to strong electron-phonon interaction in Pt, since electrons in the Pt contact are exposed to phonon heat currents in the longitudinal configuration owing to the direct contact between the Pt and the heat bath. This difference between the longitudinal and transverse configurations can explain the fact that the sign of the spin current generated by the longitudinal SSE at the YIG/Pt interface (Fig. 4) is opposite to that by the transverse SSE (Fig. 7).[7]

## V. SUMMARY

In the present study, we investigated the spin-Seebeck effect (SSE) in ferromagnetic metals and ferrimagnetic insulators by means of the inverse spin-Hall effect (ISHE) in paramagnetic metals in the longitudinal and transverse configurations. The longitudinal configuration, in which a spin current parallel to a temperature gradient is measured, consists of a simple and straightforward structure, enabling us to easily investigate the SSE in magnetic insulators [e.g. $\mathrm{Y_3Fe_5O_{12}}$ (YIG)]. The transverse configuration, in which a spin current flowing perpendicular to a temperature gradient is measured, has been used for measuring the SSE both in metals (e.g. $\mathrm{Ni_{81}Fe_{19}}$) and insulators [e.g. $\mathrm{LaY_2Fe_5O_{12}}$ (La:YIG)].

The common mechanism for all the observed SSE phenomena appears to be thermal non-equilibrium distributions between magnons in a ferromagnetic layer and electrons in an attached paramagnetic layer. We described this non-equilibrium state in terms of an effective magnon temperature $T_m^*$ in the ferromagnet and an electron temperature $T_e^*$ in the paramagnet; the spin current generated by the SSE in the paramagnet is proportional to $T_m^* - T_e^*$. At the ferromagnet/paramagnet interface, $T_m^* - T_e^*$ is generated by the magnon- and/or phonon-mediated processes. The experimental results show that, in the $\mathrm{Ni_{81}Fe_{19}/Pt}$ sample, the SSE is dominated by the phonon-mediated process, while both magnon and phonon propagations appear to be important in the YIG/Pt and La:YIG samples. In the transverse configuration, the long-range magnon and phonon propagations through a temperature gradient are responsible for the close to linear dependence of the ISHE voltage induced by the SSE and the sign reversal of the voltage between the lower- and higher-temperature regions of the ferromagnets. The phonon-mediated process also gives rise to the giant enhancement of the SSE signals at low temperatures in single-crystalline samples. We anticipate that this systematic information on the SSE will invigorate spintronics and spin-caloritronics researches.

## ACKNOWLEDGMENTS

The authors thank S. Takahashi, J. Ieda, J. Ohe, W. Koshibae, K. Harii, and A. Kirihara for valuable discussions. This work was supported by CREST-JST "Creation of Nanosystems with Novel Functions through Process Integration", Japan, a Grant-in-Aid for Scientific Research A (21244058) from MEXT, Japan, the global COE for the "Materials Integration International Center of Education and Research" from MEXT, Japan, National Natural Science Foundation of China (11004036), the FOM Foundation, EU-ICT-7 "MACALO", and DFG Priority Programme 1538 "Spin-Caloric Transport".

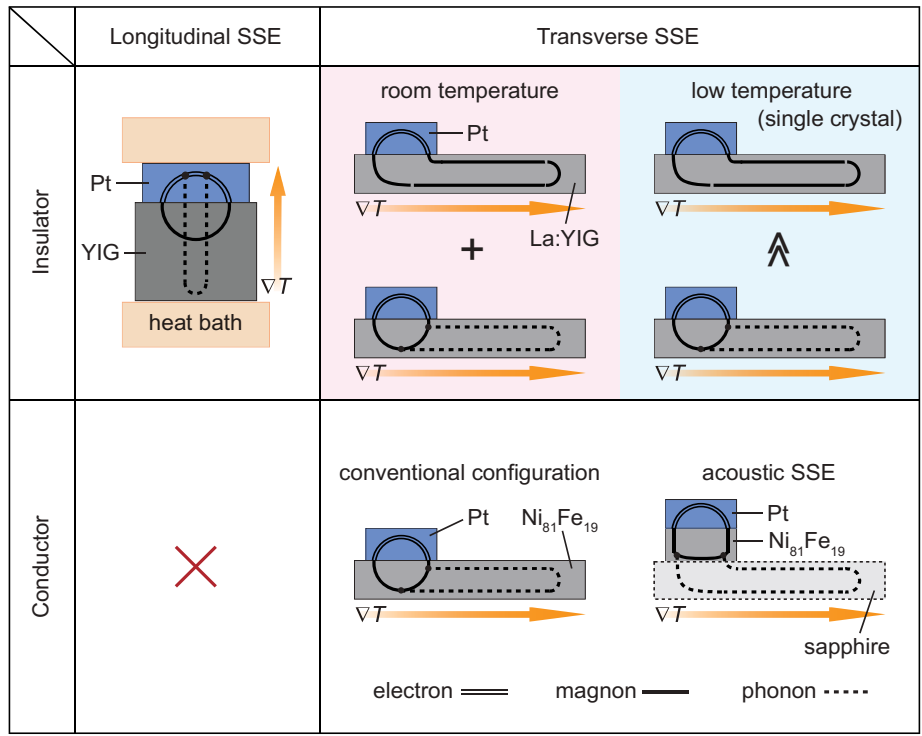

FIG. 13: Feynman diagrams representing the dominant contributions to the longitudinal and transverse SSEs in the samples used in the present study.[5,7,8,10,15] The double lines, bold lines, and dotted lines represent electron spin-density propagators, magnon propagators, and phonon propagators, respectively.